\documentclass[conference]{IEEEtran}
\IEEEoverridecommandlockouts

\usepackage{cite}
\usepackage{amsmath, amssymb, amsfonts}
\usepackage{textcomp}
\usepackage{microtype}
\usepackage[ruled,vlined]{algorithm2e}
\usepackage{algorithmic}
\usepackage{graphicx} \graphicspath{{figures/}}
\usepackage{xcolor} \definecolor{LightCyan}{rgb}{0.88,1.0,1.0}
\usepackage{booktabs,multicol, multirow, colortbl}
\usepackage{balance}
\usepackage{authblk}

\def\BibTeX{{\rm B\kern-.05em{\sc i\kern-.025em b}\kern-.08em
    T\kern-.1667em\lower.7ex\hbox{E}\kern-.125emX}}

\title{Analysis of Scoliosis From Spinal X-Ray Images}

\author[1,2]{Abdullah-Al-Zubaer Imran}
\author[2]{Chao Huang}
\author[2]{Hui Tang}
\author[2]{Wei Fan}
\author[3]{Kenneth M.C. Cheung}
\author[3]{Michael To}
\author[2]{\authorcr Zhen Qian}
\author[1,4]{Demetri Terzopoulos}
\affil[1]{University of California, Los Angeles, California, USA}
\affil[2]{Tencent Hippocrates Research Lab, Palo Alto, California, USA}
\affil[3]{The University of Hong Kong, Hong Kong, China}
\affil[4]{VoxelCloud, Inc., Los Angeles, California, USA}
    
\begin{document}

\maketitle

\begin{abstract}
 Scoliosis is a congenital disease in which the spine is deformed from its normal shape. Measurement of scoliosis requires labeling and identification of vertebrae in the spine. Spine radiographs are the most cost-effective and accessible modality for imaging the spine. Reliable and accurate vertebrae segmentation in spine radiographs is crucial in image-guided spinal assessment, disease diagnosis, and treatment planning. Conventional assessments rely on tedious and time-consuming manual measurement, which is subject to inter-observer variability. A fully automatic method that can accurately identify and segment the associated vertebrae is unavailable in the literature. Leveraging a carefully-adjusted U-Net model with progressive side outputs, we propose an end-to-end segmentation model that provides a fully automatic and reliable segmentation of the vertebrae associated with scoliosis measurement. Our experimental results from a set of anterior-posterior spine X-Ray images indicate that our model, which achieves an average Dice score of 0.993, promises to be an effective tool in the identification and labeling of spinal vertebrae, eventually helping doctors in the reliable estimation of scoliosis. Moreover, estimation of Cobb angles from the segmented vertebrae further demonstrates the effectiveness of our model.
\end{abstract}


\section{Introduction}

\begin{figure*}
    \centering
         \includegraphics[width=\linewidth, trim={0cm, 19.75cm, 2.5cm, 0cm},clip]{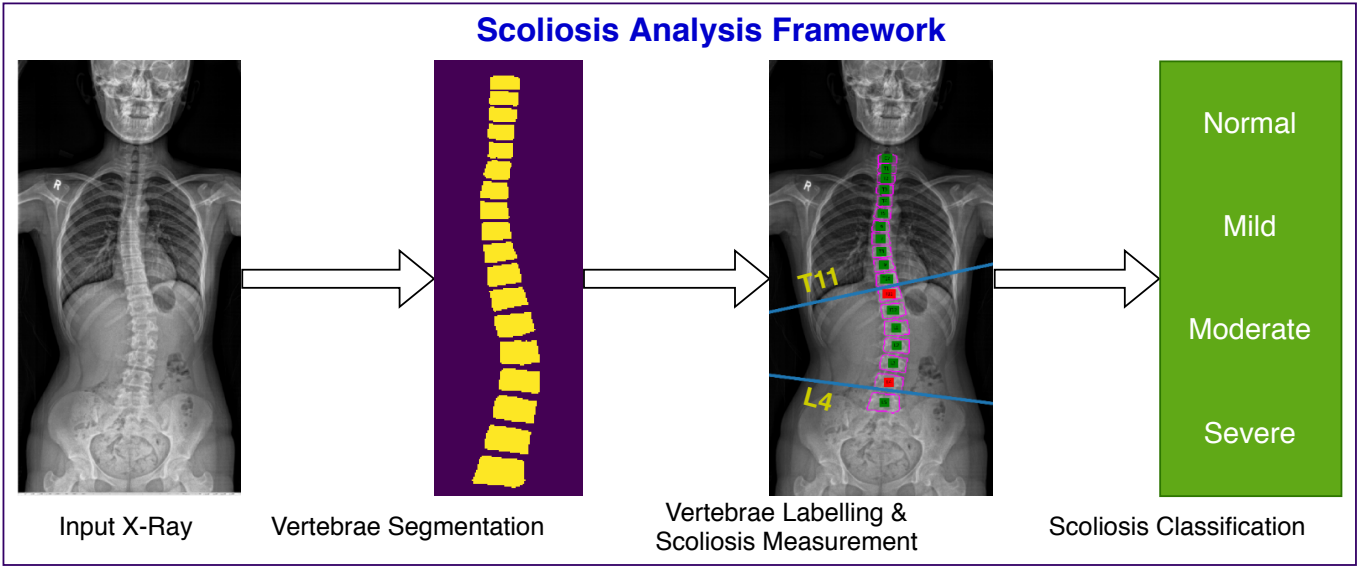}
    \caption{Overview of our framework for calculating the Cobb angle in a spine X-Ray through segmentation, labeling, and identification of the relevant vertebrae. After determining the most tilted vertebrae above and below the apex, tangents are drawn by extending the upper edge of the upper vertebra and lower edge of the lower vertebra. From these tangents, the Cobb angles are calculated and the scoliosis can be classified.}
    \label{fig:framework}
\end{figure*}

Scoliosis is an abnormal condition defined by spinal curvature towards the left or right. Early detection is key and, when accurate, it can lead to better treatment planning \cite{weinstein2008adolescent}. Radiography (X-Ray) is the preferred imaging technique for clinical analysis and measurement of scoliosis as it is highly available, inexpensive, and yields quick results. Conventional spine image analysis tasks involve tedious manual labor with hand-crafted feature extraction for the measurement of scoliosis. Cobb angle, the standard quantification of scoliosis is estimated by calculating the angle between the two tangents of the upper and lower end plates of the upper and lower vertebrae. A person with a $10^{\circ}$ or greater Cobb angle is usually considered for scoliosis diagnosis \cite{kim2010scoliosis}. Fig.~\ref{fig:framework} illustrates the calculation of the Cobb angle and the labeling of relevant vertebrae in an X-Ray image. 

Conventionally, measurement and assessment, which requires the identification and labeling of specific vertebral structures, is manually performed by clinicians. However, the manual measurement of scoliosis faces several difficulties. First, large anatomical variation between patients and low tissue contrast in spinal X-Ray images make it challenging to accurately and reliably assess the severity of scoliosis \cite{wu2017automatic}, and effects on the spine and body as a whole, as well as on individual vertebra, pose extra difficulty in the quantification of scoliosis \cite{kawchuk1997scoliosis}. Second, measurement error is prevalent in the routine clinical assessment of scoliosis due to instrumentation, vertebral rotation, and patient positioning \cite{kim2010scoliosis}, and $5^{\circ}$--$10^{\circ}$ intra- or greater inter-observer variation has commonly been reported in measuring the Cobb angle \cite{beauchamp1993diurnal, pruijs1994variation}.

Therefore, an automatic technique for the accurate measurement of scoliosis is desirable. Our specific contributions in this paper are the following: 
\begin{enumerate}
    \item A fully automatic and efficient pipeline for the measurement and analysis of scoliosis.
    \item A novel segmentation network for accurately segmenting vertebrae from spine X-Ray images.
    \item Fully automatic and accurate identification and labeling of individual vertebrae merely based on binary segmentation.
    \item Accurate diagnostic classification of the severity of scoliosis, which is crucial for treatment planning.
\end{enumerate}

\section{Related Work}

While several methods for vertebrae segmentation and scoliosis measurement are available, this approach is still under-explored in the literature. Existing vertebrae segmentation methods rely on manual interaction \cite{malgorzata2019semi}, hand-crafted feature engineering limited to customized parameters \cite{taghizadeh2019automated, anitha2012automatic}, follow patch-based approaches that lose full spatial context \cite{qadri2019vertebrae, horng2019cobb}, are limited in scope and fail to consider all the required vertebrae at a time \cite{lessmann2019iterative}, etc. For Cobb angle estimation, a minimum bounding rectangle was used for the patch-wise segmented vertebrae \cite{horng2019cobb}, an approach that relies on pre-processing steps including spinal region isolation and vertebrae detection. Kusuma et al. \cite{Kusuma2017Determination} proposed a K-means and curve-fitting approach for Cobb angle measurement that requires a set of pre-processing steps \cite{Kusuma2017Determination}. Other Cobb angle estimation methods have been proposed based on directly finding vertebrae corners as a form of regression task \cite{wu2018automated, sun2017direct, imran2019bipartite, wu2017automatic}. Although promising, these supervised methods are less viable for clinical applications because of low accuracy, due to the loss of fine details in the process, and the lack of explainability.

As a departure from prior segmentation-based methods, our model is fully automatic, involving no manual intervention end-to-end, and eschews any kind of pre-processing or post-processing steps.

\begin{figure*}
    \centering
    \includegraphics[width=\linewidth, trim={0cm, 6cm, 0cm, 6cm}, clip]{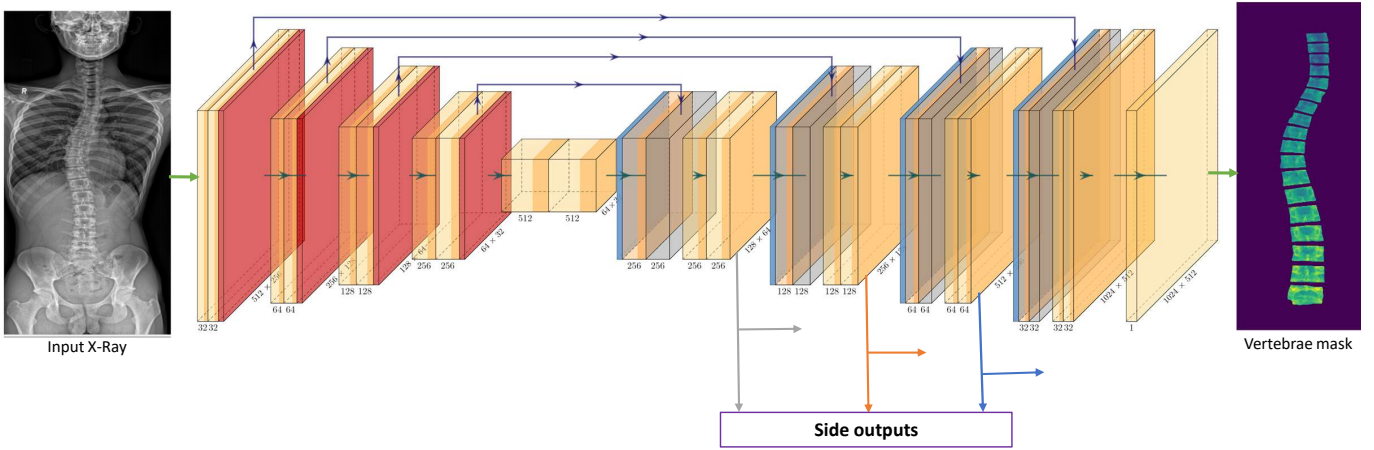}
    \caption{Architecture of our segmentation network (Progressive U-Net): Side outputs at three different stages of the decoder are generated and progressively added to the next stage side-output. The output from the third side-output is added to the last stage before the final convolution to generate the final segmentation output.}
    \label{fig:seg_net}
\end{figure*}

\section{Methods}

\subsection{Vertebrae Segmentation and Labeling}

We perform binary segmentation of the spine with a well-distinguishable number $n$ of vertebrae relevant to scoliosis analysis. To formulate the problem, we assume an unknown data distribution $p(X,Y)$ over images $X$ and vertebrae segmentation labels $Y$. The model has access to the labeled training set $\mathcal{D}_{(x,y)}$ sampled i.i.d. from $p(X, Y)$. As illustrated in Algorithm~\ref{alg:train}, the segmentation prediction network $\mathcal{F}_\phi$ is trained with a set of learnable parameters $\phi$. We specify the objective as $\min_{\phi_\mathcal{F}} \mathcal{L}_{(y, \hat{y})}$, where $y$ is the reference vertebrae mask and $\hat{y}$ is the model prediction in each of the training iterations.

\begin{algorithm}[t]
\caption{Training for vertebrae segmentation from spine X-Ray images.} {\bf Input:} X-Ray images and reference vertebra masks.\\
{\bf Output:} Predicted vertebra masks.

\label{alg:train}
\begin{algorithmic}
\REQUIRE 
\STATE Training data $x, y \in \mathcal{D}$ including spine X-Ray images $x$ and reference vertebra segmentation masks $y$
\STATE Model architecture $\mathcal{F_\phi}$ with learnable parameters $\phi$
\vspace{4pt}
\FOR{each step over $\mathcal{D}$}
\STATE Sample minibatch $\mathcal{M}: $ $x_{(i)} \sim p_{{\mathcal{D}_{(x)}}}$

\vspace{4pt}
\STATE Compute model outputs for the minibatch:\\
$\hat{y}_{(i)} \leftarrow \mathcal{F}_{(\phi)}(x)$ 

\STATE Calculate loss $\mathcal{L}_{(y,\hat{y})}$ for the model predictions
\vspace{4pt}
\STATE Update the model $\mathcal{F}$ along its gradient
\begin{align*}
    \nabla_{\phi_{\mathcal{F}}} &\frac{1}{|\mathcal{M}|}
    \sum_{i\in\mathcal{M}}
    \left[
    \mathcal{L}_{\mathcal{F}_{\left(y_{(i)}, \hat{y}_{(i)}\right)}}
    \right]
\end{align*}
\ENDFOR
\end{algorithmic}
\end{algorithm}

 Following the progressive dense V-net model \cite{imran2018automatic, imran2019fast}, we propose a progressive U-Net with some careful adjustments in the U-Net \cite{ronneberger2015u}. As shown in Fig.~\ref{fig:seg_net}, our model has an encoder and a decoder with skip connections. In each encoder layer, two $3\times3$ convolutions are followed by instance normalization, ReLU activation, and a $2\times2$ max-pooling. A dropout is applied in every encoder and decoder stage of the network. We generate side-outputs in every stage of the decoder. Progressively adding one side-output to the next, the segmentation performance is improved compared to collecting the final output from the final decoder stage in a U-Net. However, one key difference with \cite{imran2018automatic} is that our model is trained without side-supervision. Only the side-outputs are generated and added progressively, yielding an improved segmentation at the final output. A convolution operation is performed to generate the side-output from each decoder stage. The progressive side-outputs also ensure that micro-structure is not lost from any level of the decoder through the convolutional operations. We generate side outputs at $x/8$, $x/4$, and $x/2$ resolutions before the final output at $x$ resolution. Therefore, the side output at resolution $x/8$ is added to the next decoder stage, and so on.     

\begin{algorithm}[t]
\caption{Cobb angle calculation}
{\bf Input:} Vertebra mask $\hat{y}$.\\
{\bf Output:} Cobb angle $\theta$.

\label{alg:cobb}
\begin{algorithmic}
\STATE From the predicted mask $\hat{y}$, get all the contours
\FOR{each contour in contours}
\IF{Number of pixels $< a$} 
\STATE //to remove any noisy patches
\STATE Remove contour 
\ENDIF
\ENDFOR
\STATE This will give $n$ contours of well-separated vertebrae 
\vspace{4pt}
\STATE Extract four corner points for the contour 
\STATE Order the corners from bottom to top by comparing the coordinates of the extracted $4n$ corner points  
\vspace{4pt}
\STATE Find the two vertebrae (upper and lower) with at least 2 vertebrae gap between them
\vspace{4pt}
\STATE Calculate the Cobb angle,
    $\theta = \left|\tan^{-1}\left(\frac{m_u - m_l}{1 + m_u m_l}\right)\right|$, where $m_u$ and $m_l$ are the upper and lower vertebrae slopes.
\end{algorithmic}
\end{algorithm}

\begin{figure*}
\centering
 \resizebox{\linewidth}{!}{%
  \begin{tabular}{c c c}
    {\Huge Raw} & {\Huge Mask} & {\Huge Measurement}\\
    \smallskip\\
     \includegraphics[width=\linewidth, trim={0cm, 1cm, 9cm, 1cm},clip]{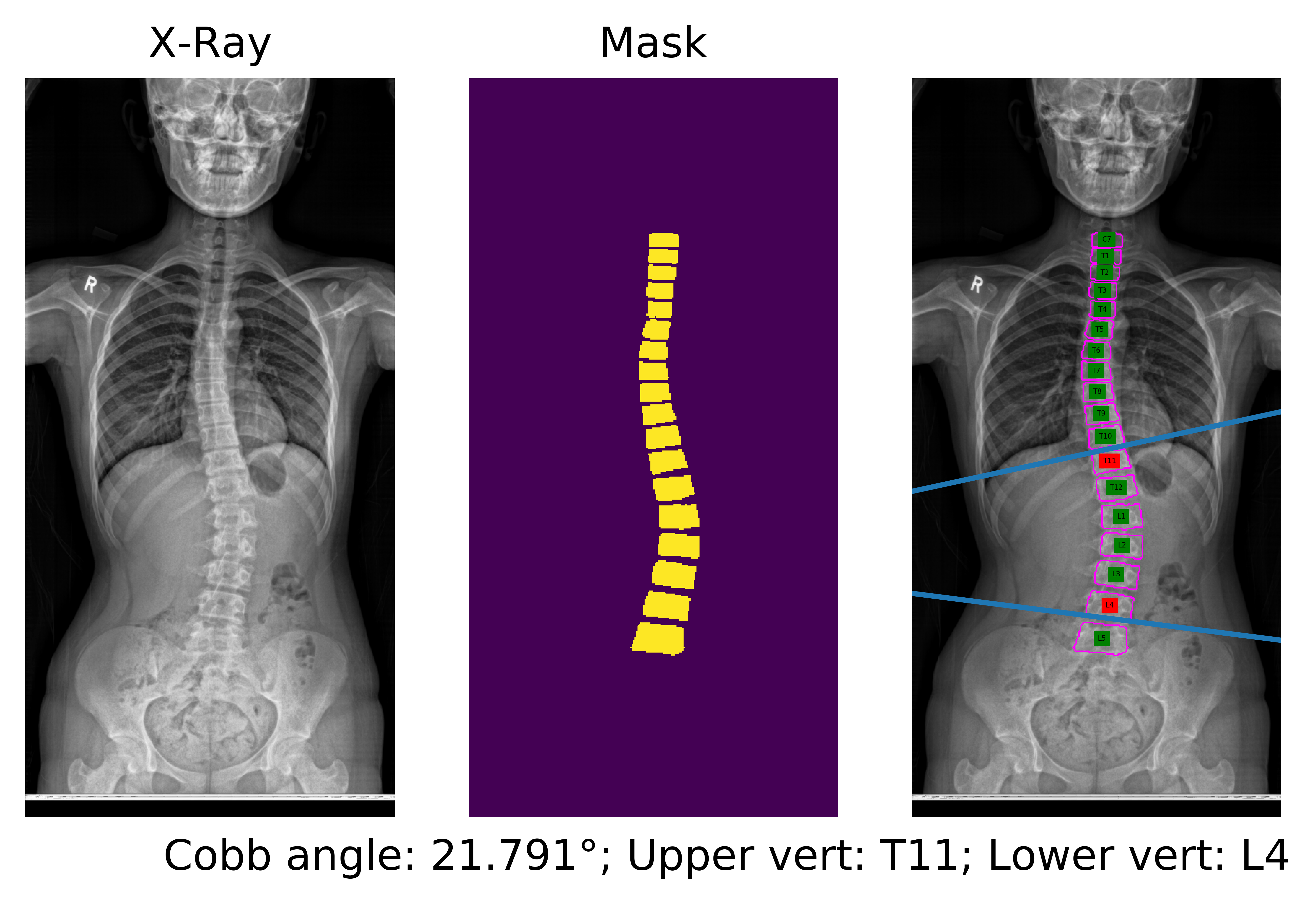}
     &
     \includegraphics[width=\linewidth, trim={4.5cm, 1cm, 4.5cm, 1cm},clip]{cobb_result1}
     &
     \includegraphics[width=\linewidth, trim={9cm, 1cm, 0cm, 1cm},clip]{cobb_result1}
     \\
     &\includegraphics[width=\linewidth, trim={0cm, 0cm, 0cm, 8.25cm},clip]{cobb_result1}\\
     \smallskip\\
     \includegraphics[width=\linewidth, trim={0cm, 1cm, 9cm, 1cm},clip]{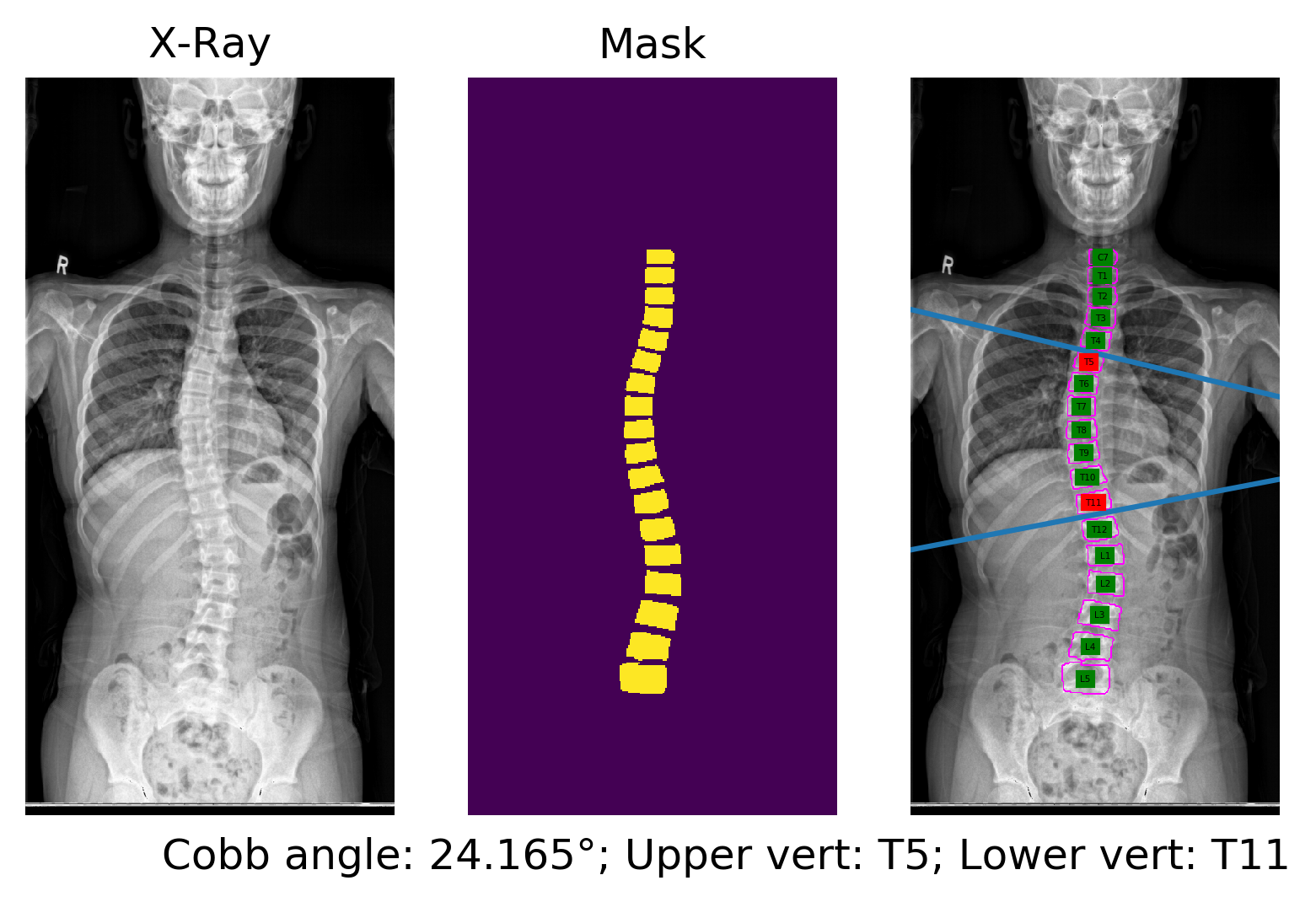}
     &
     \includegraphics[width=\linewidth, trim={4.5cm, 1cm, 4.5cm, 1cm},clip]{cobb_result2}
     &
     \includegraphics[width=\linewidth, trim={9cm, 1cm, 0cm, 1cm},clip]{cobb_result2}
     \\
     &\includegraphics[width=\linewidth, trim={0cm, 0cm, 0cm, 8.25cm},clip]{cobb_result2}\\
    \end{tabular} }
    \caption{From input X-ray image (left), to segmentation mask prediction (middle), to vertebrae identification and scoliosis measurement (right) in our pipeline.}
    \label{fig:xray-to-cobb}
\end{figure*}

\subsection{Measurement of Scoliosis}

Our pipeline makes use of the vertebrae segmentation in estimating Cobb angles. Algorithm \ref{alg:cobb} automatically calculates the Cobb angle by analyzing the contours from the segmented mask. When well-separated from others, each of the contours represents a vertebra relevant to the measurement of scoliosis. To verify if a contour is actually associated to a relevant vertebra, we impose a minimum size on the number of contour pixels ($a$). After the extraction and ordering of $4n$ corners, the most tilted upper vertebra and the most tilted lower vertebra are determined from the $n$ relevant vertebrae (Fig.~\ref{fig:xray-to-cobb}). Then the Cobb angle is calculated from the slopes of the upper edge of the upper vertebra and the lower edge of the lower vertebra.

Moreover, the severity of scoliosis can be categorized and appropriate treatment planning is performed depending on the calculated Cobb angle from the spine X-Ray of a patient. In our pipeline, we therefore perform an automatic diagnostic classification following the clinically recognized scoliosis severity classes, as shown in Table~\ref{tab:scoliosis_class}. Active treatment is typically not needed when it is mild and rigid braces can stop the progression of scoliosis when it is in moderate stage. Surgery is the last resort for severe cases, but it can be delayed for the adolescent period \cite{yang2016early}.     

\begin{table}[t]
    \centering
    \caption{Clinically accepted classification and treatment planning for adolescent scoliosis based on measured Cobb angles}
    \label{tab:scoliosis_class}
    \medskip
    \begin{tabular}{l l l}
        \toprule
         Cobb Angle $\theta$ ($^{\circ}$) & Severity & Treatment Recommendation \\
         \midrule
        $\theta < 10^{\circ}$ & normal & ---\\
         $10^{\circ} < \theta < 25^{\circ}$ & mild & Check in every 2 years\\
         $25^{\circ} < \theta < 45^{\circ}$ & moderate & Wear a brace for 16--23 hours/day\\
         $45^{\circ} < \theta$ & severe & Revision surgery in 20--30 years\\
         \bottomrule
    \end{tabular}
\end{table}

\section{Experimental Evaluation}

\subsection{Implementation Details} 

{\bf Data:} We use a dataset of 100 high-resolution spine X-Ray images of children with evidence of scoliosis to various extents. The dataset contains manual annotation by experts of 18 relevant vertebrae (cervical C7, thoracic T1--T12, lumbar L1--L5). We split the dataset into training (80), testing (15), and validation (5) sets. {\bf Baselines:} As baselines, we use a regular U-Net model with a choice of binary cross-entropy (XE) and Dice as loss function. For simplicity, we denote the models as UD (UNet with Dice loss), UX (UNet with XE loss), PUD (Progressive UNet with Dice loss), and PUX (Progressive UNet with XE loss). {\bf Training:} The models are trained on the training set while their performances were evaluated on the testing set. The validation set is used for hyper-parameter tuning and model selection. {\bf Inputs:} All the images are resized and normalized to $1024\times512\times1$ before feeding them to the network. {\bf Hyperparameters:} We use the Adam optimizer with adaptive learning rate starting with an initial rate of 0.01 and decreasing 10 times after every 20 epochs. We apply dropout with a rate of 0.25. {\bf Machine Configuration:} We implemented Algorithm~\ref{alg:train} in TensorFlow running on a Tesla P40 GPU in a system with a 64-bit Intel(R) Xeon(R) 440G CPU. {\bf Segmentation Evaluation:} For segmentation evaluation, along with qualitative visualization of masks and edges, we use the Dice index (DI), structural similarity index (SSIM), average Hausdorff distance (HD), and F1 score (F1). {\bf Scoliosis Evaluation:} For the evaluation of scoliosis, we measure Cobb angles, the indices of upper and lower tilted vertebrae, and severity classification. Since the expert annotations include only the vertebrae labels for segmentation reference, we follow the same scoliosis measurement procedure for both the reference measurements and for our progressive U-Net-based approach.

\subsection{Segmentation Results}

\begin{figure}
\centering
 \resizebox{\linewidth}{!}{%
  \begin{tabular}{cccccc}
    \includegraphics[width=0.2\linewidth, trim={4cm 1cm 3cm 1cm},clip]{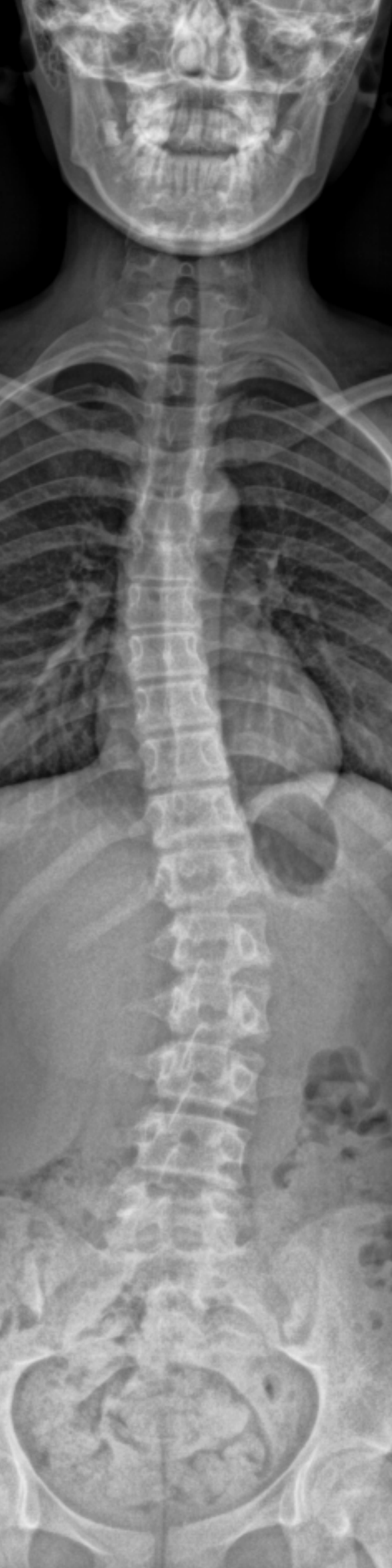}
    &
     \includegraphics[width=0.2\linewidth, trim={4cm 1cm 3cm 1cm},clip]{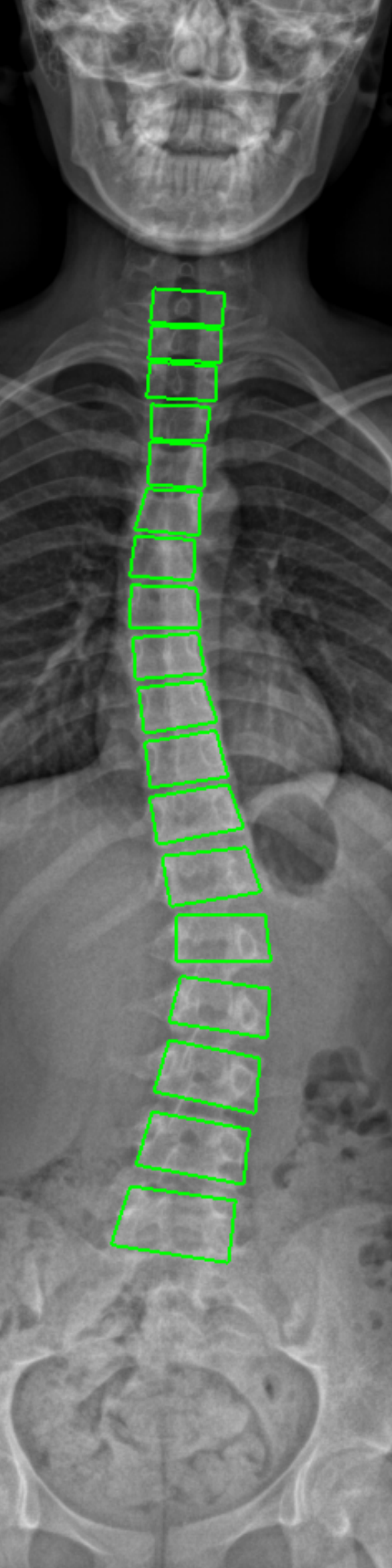}
    &
    \includegraphics[width=0.2\linewidth, trim={4cm 1cm 3cm 1cm},clip]{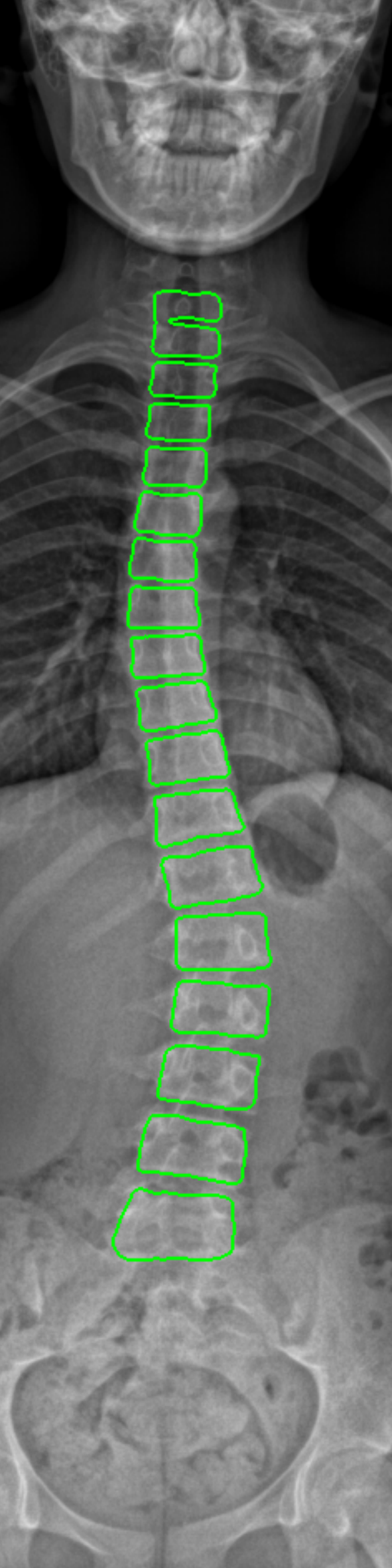}
    &
    \includegraphics[width=0.2\linewidth, trim={4cm 1cm 3cm 1cm},clip]{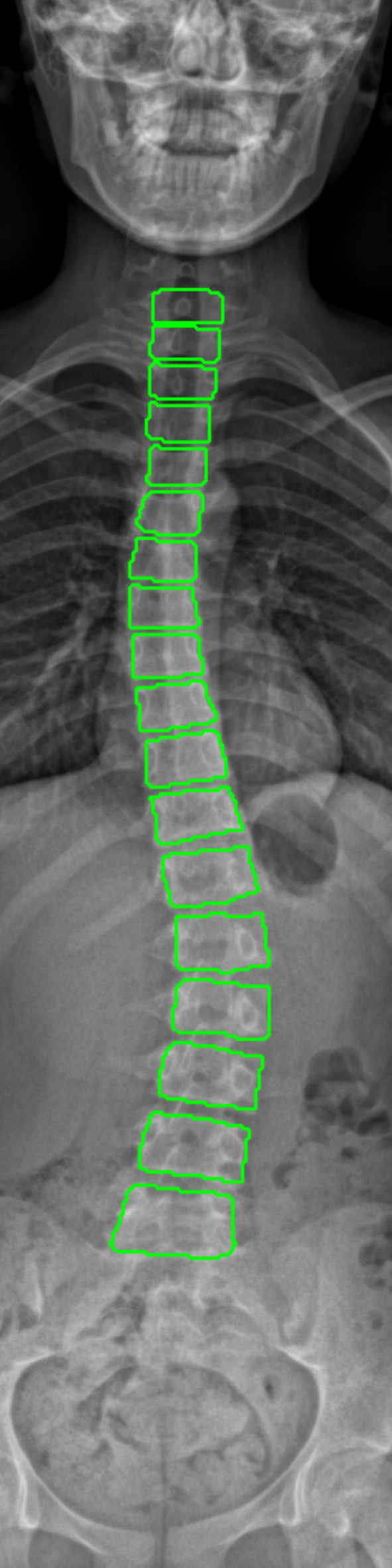}
    &
    \includegraphics[width=0.2\linewidth, trim={4cm 1cm 3cm 1cm},clip]{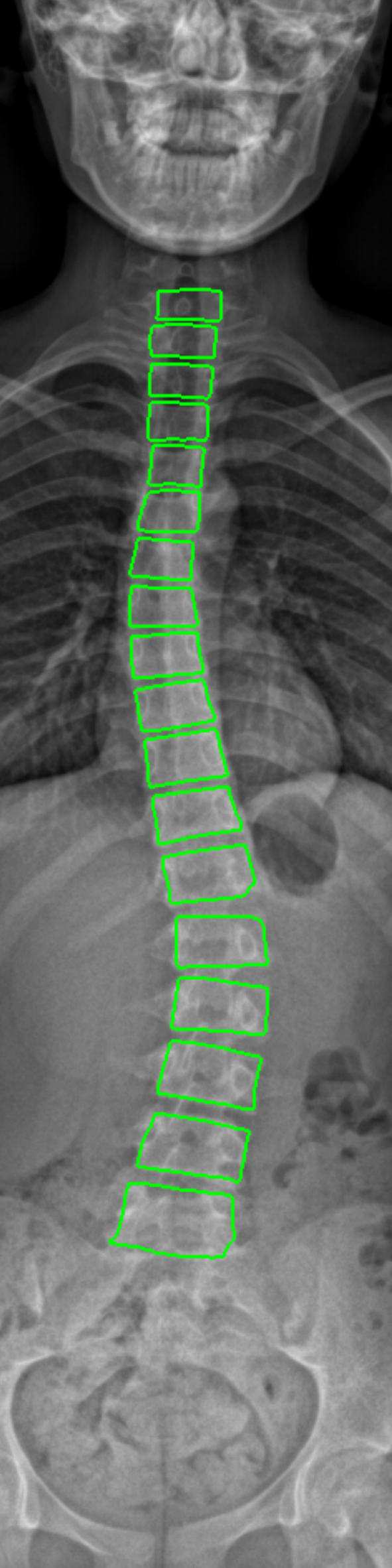}
    &
    \includegraphics[width=0.2\linewidth, trim={4cm 1cm 3cm 1cm},clip]{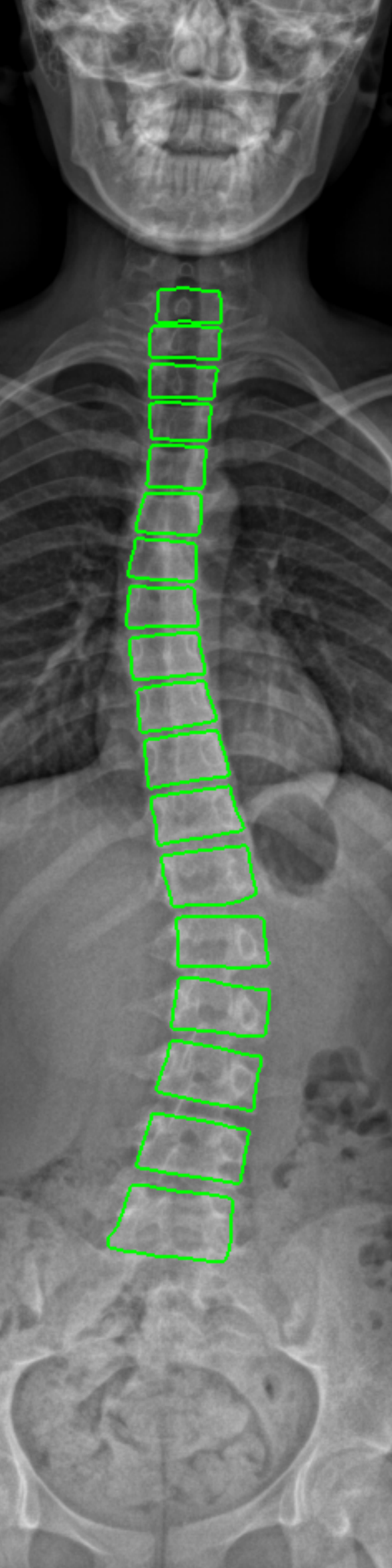}
    \\
    {\Large Raw}&{\Large GT}&{\Large UD}&{\Large\bf PUD}&{\Large UX}&{\Large\bf PUX}\\
    \end{tabular}
    }
    \caption{Boundary visualization of the predicted vertebrae masks in a spine X-Ray shows consistent improvement by our model over all other models.}
    \label{fig:edge}
\end{figure}

\begin{figure}
\centering
 \resizebox{\linewidth}{!}{%
  \begin{tabular}{cccccc}
    \includegraphics[width=0.2\linewidth, trim={4cm 1cm 3cm 1cm},clip]{raw_image}
    &
     \includegraphics[width=0.2\linewidth, trim={4cm 1cm 3cm 1cm},clip]{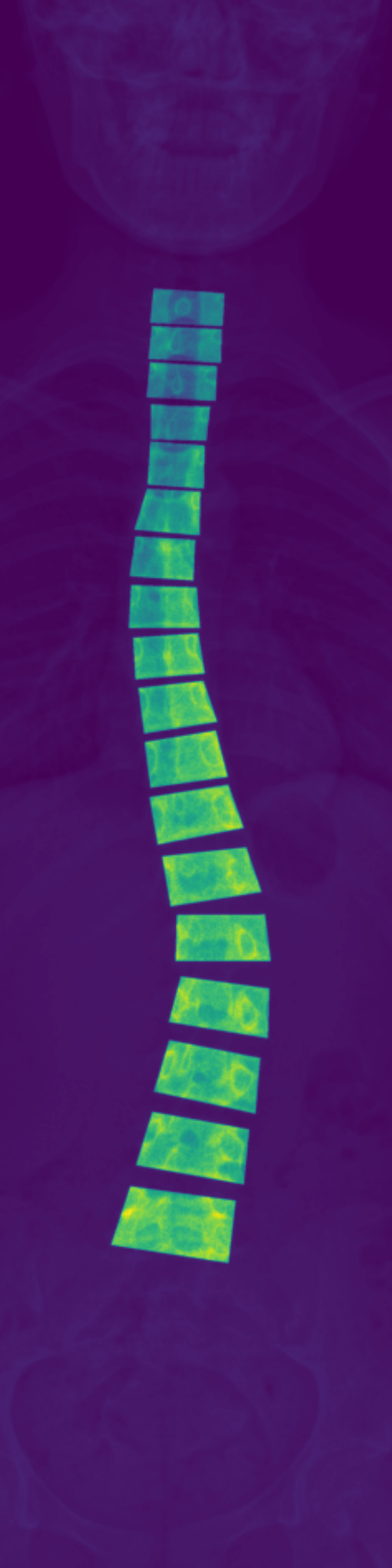}
    &
    \includegraphics[width=0.2\linewidth, trim={4cm 1cm 3cm 1cm},clip]{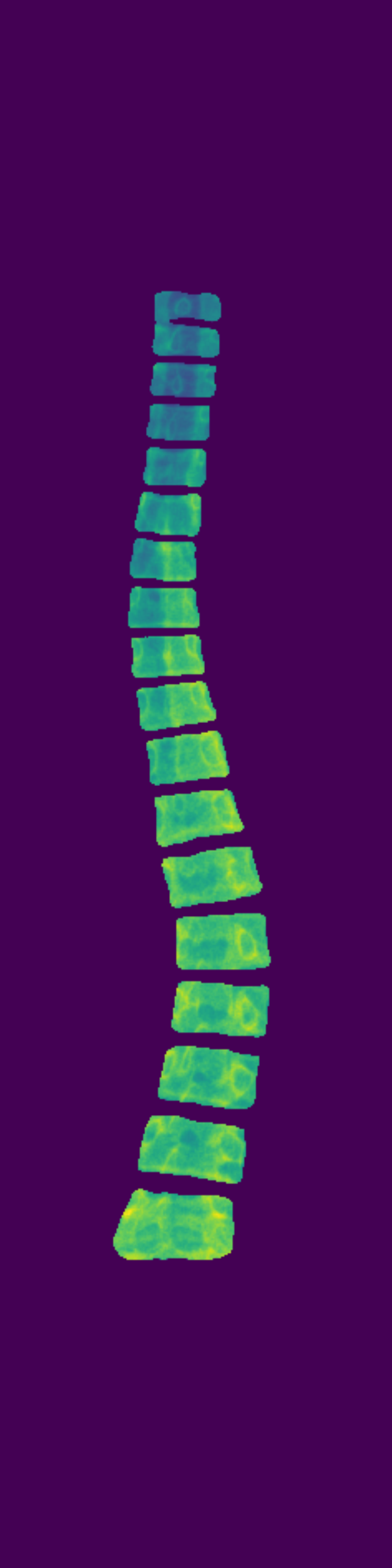}
    &
    \includegraphics[width=0.2\linewidth, trim={4cm 1cm 3cm 1cm},clip]{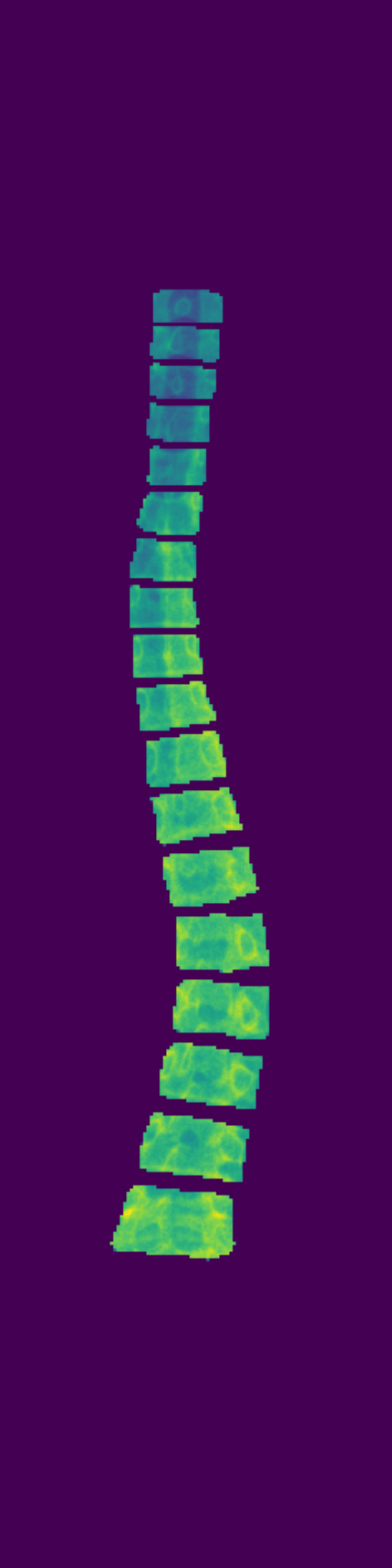}
    &
    \includegraphics[width=0.2\linewidth, trim={4cm 1cm 3cm 1cm},clip]{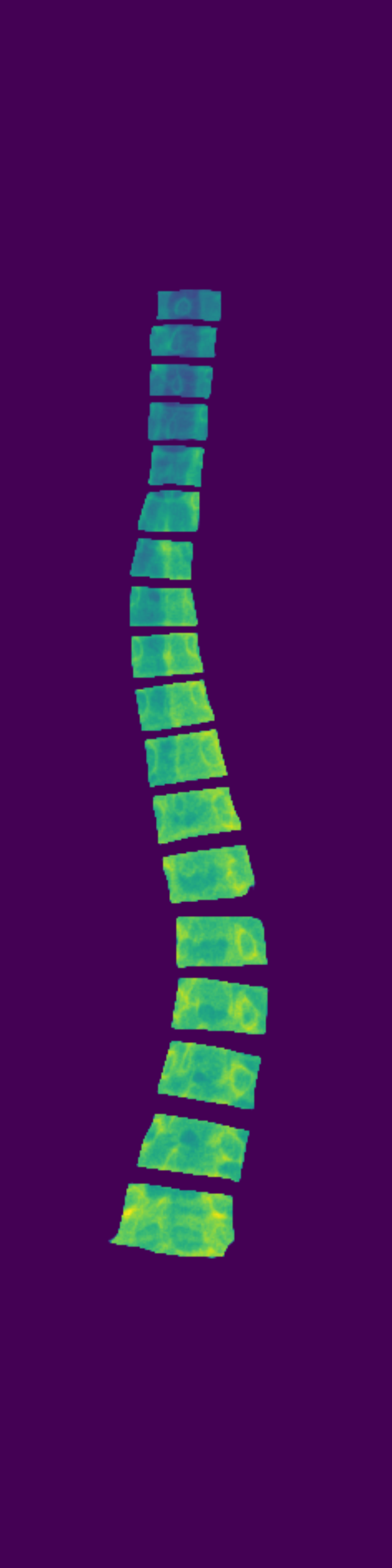}
    &
    \includegraphics[width=0.2\linewidth, trim={4cm 1cm 3cm 1cm},clip]{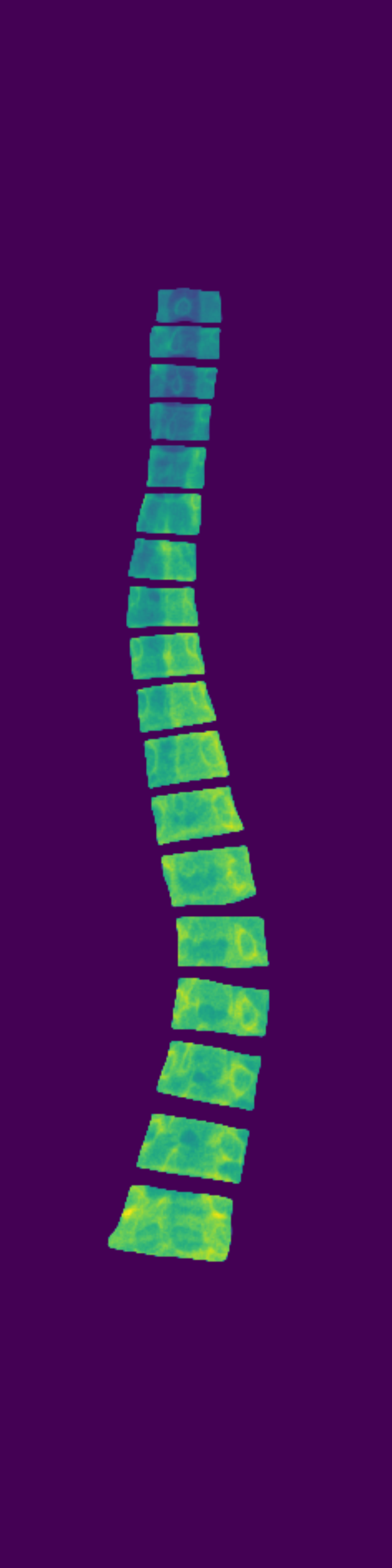}
    \\
    {\Large Raw}&{\Large GT}&{\Large UD}&{\Large\bf PUD}&{\Large UX}&{\Large\bf PUX} \\
    \end{tabular}
    }
     \caption{Visualization (zoomed) of the predicted vertebrae masks in a spine X-Ray shows consistent improvement by our model over all other models.}
    \label{fig:mask}
\end{figure}

\begin{figure}
    \centering
    \includegraphics[width=0.8\linewidth, trim={0cm 0cm 4cm 0cm}, clip]{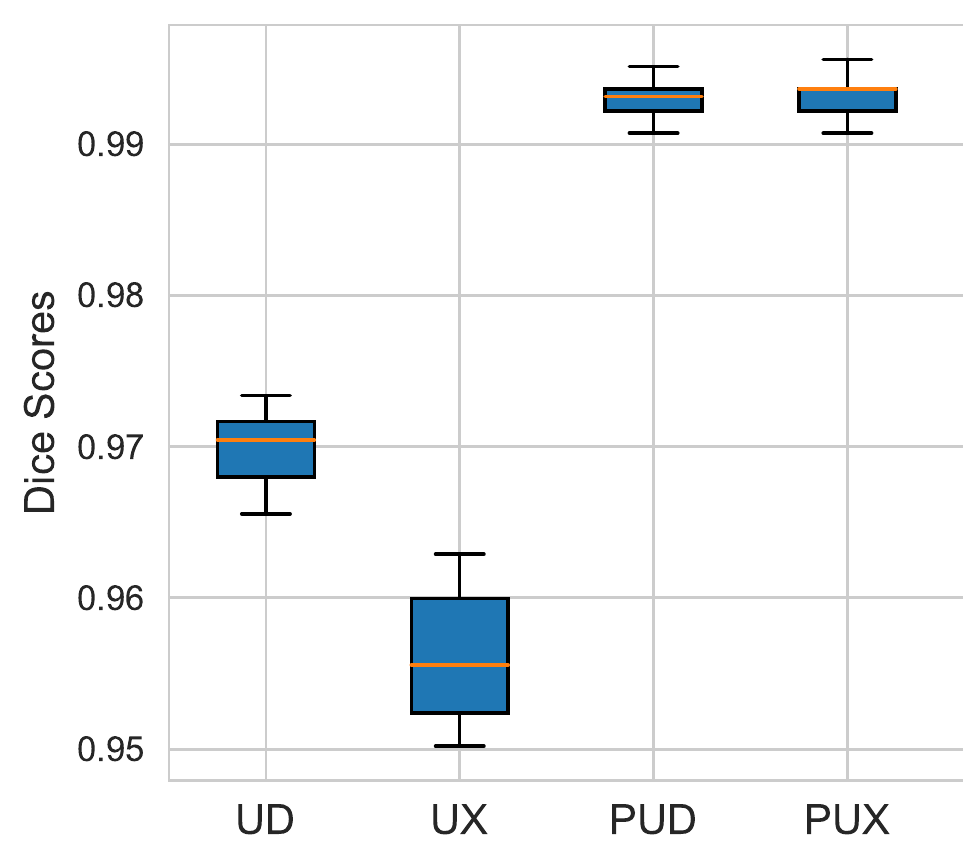}
    \caption{Whisker-Box plots of all four models showing consistent performance of our model with varying losses in segmenting 18 scoliosis-relevant vertebrae from the spine X-ray test set.}
    \label{fig:wb}
\end{figure}

Experimental results based on both qualitative and quantitative evaluations confirm the superiority of our model, which consistently provides improved segmentation with different losses (Dice and XE). Visualizations of the segmented vertebrae (Fig.~\ref{fig:edge} and Fig.~\ref{fig:mask}) depict better distinctions of the individual vertebrae merely with binary segmentation. In all four quantitative measures, our models achieve better scores than the baseline models (Table \ref{tab:seg}). The superiority of our models is further confirmed by the whisker-box plots in Fig.~\ref{fig:wb}. 
Our end-to-end vertebrae segmentation achieves a better Dice similarity score than the recently published patch-wise segmentation method \cite{horng2019cobb} (0.993 vs 0.952). While superior DI and F1 justifies the progressive addition of the side-outputs in pixel-wise predictions, better SSIM and HD depict the model's ability to learn the intrinsic shape and structure of the segmented vertebrae. 

\begin{table}
    \centering
    \caption{Performance comparison of the vertebrae segmentation models}
    \label{tab:seg}
    \begin{tabular}{lcccc}
    \toprule
         Model & DI & SSIM & HD & F1\\
         \midrule
         UD & 0.970 & 0.961 & 5.246 & 0.896\\
         UX & 0.956 & 0.955 & 6.767 & 0.868\\
         {\bf PUD} & {\bf0.993} & 0.966 & {\bf 4.597} & 0.919\\
         {\bf PUX} & {\bf 0.993} & {\bf 0.970} & 4.677 & {\bf 0.922}\\
         \bottomrule
    \end{tabular}
\end{table}

\subsection{Scoliosis Results}

For the evaluation of scoliosis, we compare the performance of our PUX model-based measurement against the reference measurement obtained by processing the expert's annotations. As reported in Table~\ref{tab:cobb}, our segmentation-based pipeline achieves very accurate Cobb angles. Good agreement is observed between our model and the reference measurement in each of the X-Rays in the test set with a mean angle difference of just 2.41 degrees, which is well below the acceptable error limit recommended by the experts \cite{cassar2002imaging}. Comparing with some of the existing Cobb angle measurement techniques, our method achieves lower measurement error than those reported in \cite{Kusuma2017Determination} and \cite{horng2019cobb}. Moreover, the categorization of scoliosis \cite{chowanska2012school} indicates 100\% diagnostic accuracy of our approach relative to the reference. 

\begin{table*}[t]
    \centering
    \caption{Performance of our method for calculating Cobb angle and scoliosis severity in the test set relative to reference measurements}
    \medskip
    \begin{tabular}{r|llrl|llrl}
    \toprule
    \multirow{2}{*}{}
    Test & \multicolumn{4}{c|}{Reference Measurement} & \multicolumn{4}{c}{Predicition of our PUX Model} \\
        ID~  & Upper Vert & Lower Vert & Cobb Angle & Severity & Upper Vert & Lower Vert & Cobb Angle & Severity\\
     \midrule
     \rowcolor{LightCyan}
     1 & T10 & L3 & $21.92^{\circ}$ & mild & T10 & L3 & $19.26^{\circ}$ & mild\\
     2 & T12 & L4 & $9.52^{\circ}$ & normal & T6 & L3 & $4.75^{\circ}$ & normal\\
     \rowcolor{LightCyan}
     3 & T11 & L3 & $13.88^{\circ}$ & mild & T11 & L3 & $14.48^{\circ}$ & mild\\
     4 & T5 & T10 & $18.78^{\circ}$ & mild & T5 & T11 & $16.16^{\circ}$ & mild\\
     \rowcolor{LightCyan}
     5 & T6 & T10 & $20.53^{\circ}$ & mild  & T10 & L4 & $20.22^{\circ}$ & mild\\
     6 & T11 & L4 & $20.38^{\circ}$ & mild & T11 & L4 & $23.35^{\circ}$ & mild\\
     \rowcolor{LightCyan}
     7 & T10 & L2 & $40.71^{\circ}$ & moderate & T11 & L3 & $41.38^{\circ}$ & moderate\\
     8 & T6 & T12 & $23.96^{\circ}$ & mild & T6 & T12 & $20.93^{\circ}$ & mild\\
     \rowcolor{LightCyan}
     9 & T5 & T11 & $21.07^{\circ}$ & mild & T6 & T11 & $23.22^{\circ}$ & mild\\
     10 & T6 & T10 & $14.81^{\circ}$ & mild & T1 & L3 & $16.10^{\circ}$ & mild\\
     \rowcolor{LightCyan}
     11 & T12 & L4 & $31.94^{\circ}$ & moderate & T12 & L4 & $28.60^{\circ}$ & moderate\\
     12 & T10 & L1 & $24.82^{\circ}$ & mild & T9 & L1 & $18.92^{\circ}$ & mild\\
     \rowcolor{LightCyan}
     13 & T12 & L3 & $15.69^{\circ}$ & mild & T10 & L3 & $14.79^{\circ}$ & mild\\
     14 & T12 & L4 & $24.25^{\circ}$ & mild & T8 & T12 & $22.72^{\circ}$ & mild\\
     \rowcolor{LightCyan}
     15 & T12 & L3 & $21.02^{\circ}$ & mild & T11 & L3 & $18.61^{\circ}$ & mild\\
     \bottomrule
    \end{tabular} 
    \label{tab:cobb}
\end{table*}

\section{Conclusions}

The accurate and reliable segmentation of vertebrae is a prerequisite for the effective measurement of scoliosis. To this end, we have established a new state-of-the-art in fully automatic vertebrae segmentation in spinal X-Ray images. Our novel framework for accurately assessing scoliosis from anterior-posterior spine radiographs makes use of an end-to-end model that can accurately and reliably segments spinal vertebrae, outputting a vertebrae segmentation mask that enables the accurate measurement of scoliosis through calculation of the Cobb angle. Our pipeline promises to be an effective tool for the clinical diagnosis of scoliosis as well as for decision support in treatment planning. We envision combining the measurement of scoliosis with the training phase such that our model can make more intelligent predictions. 

\balance

\bibliographystyle{IEEEtran}
\bibliography{refs}

\end{document}